\documentclass[final,5p,times,twocolumn,number,sort&compress]{elsarticle}

\usepackage[T1]{fontenc}
\usepackage{amsmath}
\usepackage{amssymb}
\usepackage{bm}
\usepackage{graphicx}
\usepackage{caption}
\usepackage{flushend}
\usepackage{xcolor}

\usepackage[colorlinks,citecolor=blue,linkcolor=blue,urlcolor=blue]{hyperref}
\biboptions{sort&compress}
\newcommand{\GeV}{\mathrm{GeV}}

\journal{Physics Letters B}

\begin{document}

\begin{frontmatter}

\title{The Pion Gravitational Form Factors and the Trace Anomaly in QCD Factorization}

\author[lecce,nanotec]{Claudio Corian\`o}
\author[taipei]{Hsiang-nan Li}
\author[lecce]{Dario Melle}

\affiliation[lecce]{
organization={Dipartimento di Matematica e Fisica, Universit\`a del Salento and INFN Sezione di Lecce; National Center for HPC, Big Data and Quantum Computing},
addressline={Via Arnesano},
city={Lecce},
postcode={73100},
country={Italy}}

\affiliation[nanotec]{
organization={CNR Nanotec},
city={Lecce},
country={Italy}}

\affiliation[taipei]{
organization={Institute of Physics, Academia Sinica},
city={Taipei},
postcode={115},
country={Republic of China}}

\begin{abstract}
We study the pion gravitational form factor in QCD factorization, focusing on the trace-anomaly component generated by the non-Abelian \(TJJ\) vertex.  The calculation combines a Sudakov-resummation-improved pion hard kernel with the anomaly form factor suggested by momentum-space conformal field theory and by the perturbative dilaton sum rule.  Comparison with lattice QCD data shows a refined projection hierarchy: the isolated anomaly cancels in the form factor \(A_\pi(Q^2)\), while the full \(TJJ\) insertion lowers the leading-order curve at small momentum transfer squared \(Q^2\);  the anomaly is important in \(D_\pi(Q^2)\), and it gives the dominant \(TJJ\) contribution to the trace form factor.
\end{abstract}


\end{frontmatter}

\section{Introduction}

Hadron gravitational form factors (GFFs) are matrix elements of the QCD energy-momentum tensor (EMT).  They are accessed through generalized parton distributions (GPDs), exclusive reactions, and lattice QCD, and encode momentum, energy, pressure, and shear distributions~\cite{Ji:1996ek,Radyushkin:1996nd,Radyushkin:1996ru,Vanderhaeghen:1998uc,Burkert:2018bqq,Duran:2022xag}.  A pion is a particularly clean probe, whose dynamics is simplified compared to the proton case, and at large spacelike momentum transfer it can be treated in hard exclusive factorization.  Recent lattice QCD determinations of the pion GFFs~\cite{Hackett:2023pyn} therefore provide a useful benchmark for isolating short-distance mechanisms.  Perturbative analyses at large momentum transfer have also derived factorization formulae for gluon GFFs and for the quark and gluon components of hadron GFFs, including pion and proton channels~\cite{Tong:2021ctu,Tong:2022zax}.
This letter focuses on one specific contribution to the  hard scattering, related to the trace anomaly, which is part of the non-Abelian \(TJJ\) vertex to be inserted in the factorizaton formula.  The construction follows the momentum-space conformal-field-theory (CFT) organization of stress-tensor correlators~\cite{Bzowski:2013sza}, and the perturbative QCD analysis of hadron GFFs and dilaton exchange~\cite{Coriano:2024qbr,Coriano:2024pionproton}. CFT provides important insight concerning the 
decomposition of this vertex which in previous works has been investigated using the methodology of QCD in curved spacetime, where the stress energy tensor of the theory is inferred by varying the QCD action with respect to the background metric.  
The purpose of this work is to show how the anomaly-induced scalar component is projected onto the form factors \(A_\pi\), \(D_\pi\), and the trace form factor. This allows to assess the role played by the anomaly in a regime in which perturbation theory is expected to be meaningful, opening the way for future investigations and comparisons with other approaches. In this letter we are going to summarize results  which will be presented in full detail in a separate extended contribution.

\section{Pion GFFs and the Trace}
We recall that for a spin-zero hadron, such as the pion, the corresponding hadronic matrix element is expressed in terms of two form factors 
\begin{align}
\langle \pi(p_2)|T^{\mu\nu}|\pi(p_1)\rangle
&=
2P^\mu P^\nu A_\pi(t)
\nonumber\\
&\quad
+\frac{1}{2}\pi^{\mu\nu}D_\pi(t),
\label{eq:gff_letter}
\end{align}
with \(P=(p_1+p_2)/2\), \(q=p_2-p_1\), and \(t=q^2\) and 
\begin{equation}
\label{pi}
\pi^{\mu\nu}(q)=q^\mu q^\nu-g^{\mu\nu}q^2
\end{equation}
is the transverse projection operator.
 \(A_\pi(t)\) is the momentum form factor, while \(D_\pi(t)\) is the stress or \(D\)-term form factor.  The trace
\begin{equation}
\Theta_\pi(t)
\equiv
g_{\mu\nu}\langle \pi(p_2)|T^{\mu\nu}|\pi(p_1)\rangle ,
\label{eq:trace_letter}
\end{equation}
is the scalar projection of the same matrix element. 
Therefore, using Eq.~\eqref{eq:gff_letter}, the trace can be written explicitly in terms of the two form factors as
\begin{equation}
	\Theta_\pi(Q^2)
	=
	-\frac12\,q^2\,A_\pi(t)
	-\frac32\,q^2\,D_\pi(t) ,
	\label{eq:trace_explicit_letter}
\end{equation}
up to pion-mass corrections.  The explicit factor of \(Q^2\)
multiplying both form factors is the origin of the different
large-\(Q^2\) scaling of \(\Theta_\pi\) compared with \(A_\pi\) and
\(D_\pi\) separately, as discussed in Sec. \ref{interpretation}.
In massless QCD the classical EMT is traceless, but renormalization gives
\begin{equation}
T^\mu_{\ \mu}
=
\frac{\beta(g_s)}{2g_s}F^a_{\rho\sigma}F^{a\rho\sigma},
\label{eq:trace_anomaly_letter}
\end{equation}
up to quark-mass terms with the strong coupling \(g_s\) and the field tensor \(F^a_{\rho\sigma}\).    
The same form factors appear in the second Mellin moment of the pion GPD \(H_\pi\). In a common normalization one has schematically
\begin{equation}
\int dx\,x\,H_\pi(x,\xi,t)
=
A_\pi(t)+\xi^2 D_\pi(t),
\label{eq:pion_gpd_moment_letter}
\end{equation}
up to conventions for the \(D\)-term.  This relation is the bridge between hard exclusive reactions and the local EMT matrix element.  It also explains why the pion, despite the difficulty of direct measurements, remains important for future GPD analyses~\cite{Kumano:2017lhr,Savinov:2013hda,Belle:2015oin}.

For the present purpose the trace combination is especially useful.  The low-energy pion is constrained by chiral symmetry, while the large-\(|t|\) pion admits a hard-scattering treatment.  The anomaly contribution isolated below is not meant to describe all of the low-\(|t|\) dynamics; it identifies the beta-function-controlled part of the short-distance kernel and asks how that part is distributed among the three observable projections.

\section{The \(TJJ\) Hard Kernel}

For large \(Q^2=-t\), the pion EMT matrix element can be organized as a convolution of two pion distribution amplitudes with a short-distance kernel \(K^{\mu\nu}\)~\cite{Efremov:1979qk,Lepage:1980fj}.  We write schematically
\begin{equation}
K^{\mu\nu}=K^{\mu\nu}_{\rm LO}+K^{\mu\nu}_{TJJ},
\label{eq:kernel_split_letter}
\end{equation}
where \(K^{\mu\nu}_{\rm LO}\) is the leading-order (LO) contribution, and \(K^{\mu\nu}_{TJJ}\) represents the one-loop stress-tensor insertion into the two-gluon hard scattering.  The trace part of this vertex is controlled by
\begin{equation}
\mathcal{A}
=
\frac{g_s^2}{16\pi^2}\,\beta_0,
\qquad
\beta_0=\frac{11}{3}C_A-\frac{2}{3}n_f .
\label{eq:anomaly_coeff_letter}
\end{equation}
where $n_f$ denotes the number of flavours. 
The anomaly is therefore not added as a phenomenological pole, but a form factor of the renormalized stress-tensor correlator.  In the perturbative dilaton interpretation its spectral density obeys a sum rule whose integral is fixed by the anomaly coefficient~\cite{Coriano:2025sumrule}.  This is the same anomaly-pole logic developed in the CFT and in anomaly-action analyses of \(TJJ\) and related correlators~\cite{Giannotti:2008cv,Armillis:2009pq,Armillis:2010qk,Coriano:2012wp,Coriano:2014gja,Coriano:2018zdo,Coriano:2020ees,Coriano:2018bbe}. 

Denoting the spectral density of the anomalous scalar form factor isolated in the trace part of the $TJJ$ correlator as \(\rho_{\rm anom}\), we have the sum rule 
\begin{equation}
\int_0^\infty ds\,\rho_{\rm anom}(s;\{p_i^2\},m^2)
=
\mathcal{A},
\label{eq:sumrule_letter}
\end{equation}
which is mass independent. $m$ denotes the fermion mass. A dispersive test of the sum rule has been presented in \cite{Coriano:2025sumrule} and is based on an interesting pattern of cancelations of contributions from the mass-dependent spectral density. 
The area under the spectral density is fixed, while the distribution of the strength among the poles and the dispersive cut  $(s > 4 m^2)$ depends on the off-shell kinematics.

The use of momentum-space CFT is not only organizational, but has a clear interpretation in a gravitational context.  The decomposition of \(TJJ\), and more generally of tensor correlators in CFT, follows the transverse-traceless/longitudinal organization introduced in the momentum-space analysis of \cite{Bzowski:2013sza}, with modifications introduced to account for the gauge-fixing sector of QCD \cite{Coriano:2024qbr} and used in the study of anomalous graviton vertices and perturbative \(TJJ\) correlators~\cite{Coriano:2018bbe}.  In the anomaly-mediated gravitational interactions \cite{Coriano:2026vya}, the anomaly form factor couples to a diffeomorphism invariant scalar metric fluctuation, first investigated in 
\cite{2016arXiv160609220M} in the context of the dynamics of a conformalon field $\varphi$. In the QCD context, the same field defines an effective interaction which is dilaton-like $\sim \beta(g)\, \varphi F^2$. \\
In the conformal limit, the interaction mediated by the $TJJ$ vertex is described by the diffeomorphism invariant scalar component of the gravitational field decaying into two on-shell intermediate partons (gluons or quarks) \cite{Coriano:2026vya}.  \\
Concerning the motivations for introducing the conformal decomposition of \cite{Coriano:2024qbr}, we recall that the conformal Ward identities separate transverse-traceless data from longitudinal and trace sectors.  In a gauge theory the non-Abelian \(TJJ\) vertex also contains gauge-fixing and ghost contributions, so the QCD hard kernel is not obtained by a naive conformal limit.  Nevertheless, the CFT decomposition identifies the scalar anomaly form factor and clarifies why its projection is different in \(A_\pi\), \(D_\pi\) and \(\Theta_\pi\).
This distinction matters for the pion.  A scalar term in \(K^{\mu\nu}_{TJJ}\) does not multiply all GFFs by the same factor.  The trace anomaly components contributes only in the \(D\)-term and in the trace while the \(P^\mu P^\nu\) momentum, namely \(A_\pi\), coefficient is untouched by the anomaly.  The non-anomalous tensor components of the full \(TJJ\) insertion can still shift the \(A_\pi\) curve downward at low \(Q^2\).\\
This cancellation is an exact
	algebraic consequence of the tensor decomposition of the \(TJJ\)
	vertex.  The trace (anomaly) sector of the correlator is proportional
	to the single structure $\pi^{\mu\nu}(q)$ 
	\cite{Coriano:2024qbr}, as defined in \eqref{pi},
which is precisely the tensor multiplying \(D_\pi(t)\) in
	Eq.~\eqref{eq:gff_letter}.  As a result, the anomaly-proportional part of
	\(K^{\mu\nu}_{TJJ}\) cannot feed \(A_\pi\) by construction, and is
	instead captured entirely by \(D_\pi\).  This orthogonality holds
	order by order in the projection of the \(TJJ\) vertex and is
	independent of the details of the hard kernel or of the pion
	distribution amplitude used in the convolution.

\begin{figure}[t]
\centering
\includegraphics[width=0.84\columnwidth]{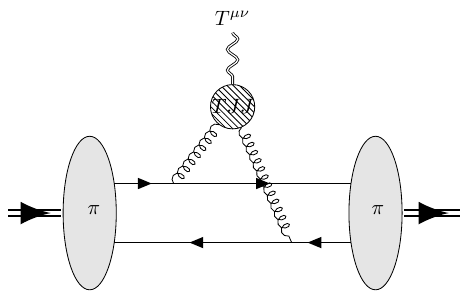}
\caption{Factorized pion GFF amplitude with the non-Abelian \(TJJ\) insertion in the hard kernel.}
\label{fig:tjj_factorization_letter}
\end{figure}

Figure~\ref{fig:tjj_factorization_letter} shows the structure used in the numerical calculation.  The initial and final pion wave functions are attached to the same hard scattering, while the stress tensor couples through the non-Abelian \(TJJ\) subgraph.  Only the anomaly-related piece of this insertion is isolated here; a complete next-to-leading-order (NLO) treatment would also contain additional radiative corrections to the pion hard kernel.

\section{Sudakov Setup and Comparison}

At finite \(Q^2\) we keep transverse-size effects in impact-parameter space and include Sudakov suppression of endpoint and large-\(b\) configurations~\cite{Botts:1989kf,Li:1992nu,Sterman:1986aj}.  The pion distribution amplitude is truncated at the first nontrivial Gegenbauer moment and combined with a Gaussian intrinsic transverse profile.  The same Sudakov-resummation-improved wave function is used for \(K^{\mu\nu}_{\rm LO}\) and for \(K^{\mu\nu}_{TJJ}\), so the comparison isolates the effect of the anomaly insertion rather than refitting nonperturbative input.

Schematically, the matrix element is factorized in the impact-parameter space as
\begin{align}
\langle \pi|T^{\mu\nu}|\pi\rangle
&=
\frac{f_\pi^2 C_F}{N_C^2}
\int dx\,dy
\int 2\pi b\,db
\nonumber\\
&\quad\times
\widetilde{\Psi}_\pi(x,b,Q)
K^{\mu\nu}(x,y,b,Q)
\widetilde{\Psi}_\pi(y,b,Q),
\label{eq:sudakov_letter}
\end{align}
where the pion wave function \(\widetilde{\Psi}_\pi\) includes the Sudakov exponent and the intrinsic transverse profile.  This treatment suppresses endpoint configurations and gives a stable way to compare the leading kernel with the anomaly-involved one.

The wave function is modeled as
\begin{align}
\widetilde{\Psi}_\pi(x,b,Q)
&=
\phi_\pi(x,\mu)\,
\exp[-S(x,b,Q)]
\nonumber\\
&\quad\times
\exp\!\left[-\frac{x(1-x)b^2}{4\beta_\pi^2}-\beta_\pi^2\frac{m_q^2}{x(1-x)}\right],
\label{eq:wave_letter}
\end{align}
with \(S\) the Sudakov exponent and $\phi_\pi$ expanded as follows
\begin{equation}
	\phi_\pi(x,\mu)=6x(1-x)\sum_{i=0}^\infty a_n(\mu)\,C_n^{3/2}(2x-1),
\end{equation}
with 
\begin{equation}
	a_n(\mu)
	=
	a_n(\mu_0)
	\left(
	\frac{\alpha_s(\mu)}{\alpha_s(\mu_0)}
	\right)^{\gamma_n},
\end{equation}
where the anomalous dimensions $\gamma_n$ are
	\begin{equation}
	\gamma_n
	=C_F	\left[	-3+ 4 \sum_{j=1}^{n+1} \frac{1}{j}-\frac{2}{(n+1)(n+2)}\right].
\end{equation}
The running coupling is evaluated at the largest hard scale in the convolution, and the transverse momentum dependence is retained in the hard gluon propagator
\begin{equation}\label{running}
	\alpha_s(\mu) =\frac{4\pi}{\beta_0\log\frac{\mu^2}{\Lambda_{QCD}^2}}.
\end{equation}
with the QCD scale $\Lambda_{\rm QCD}=0.25$ GeV is evaluated at the largest hard scale in the convolution, and the transverse momentum dependence is retained in the hard gluon propagator. 
Twist-three terms and a full Gegenbauer expansion are not included, since the aim is to isolate the channel dependence of the anomaly insertion rather than to perform a global fit.

\begin{figure*}[t]
\centering
\includegraphics[width=0.47\textwidth]{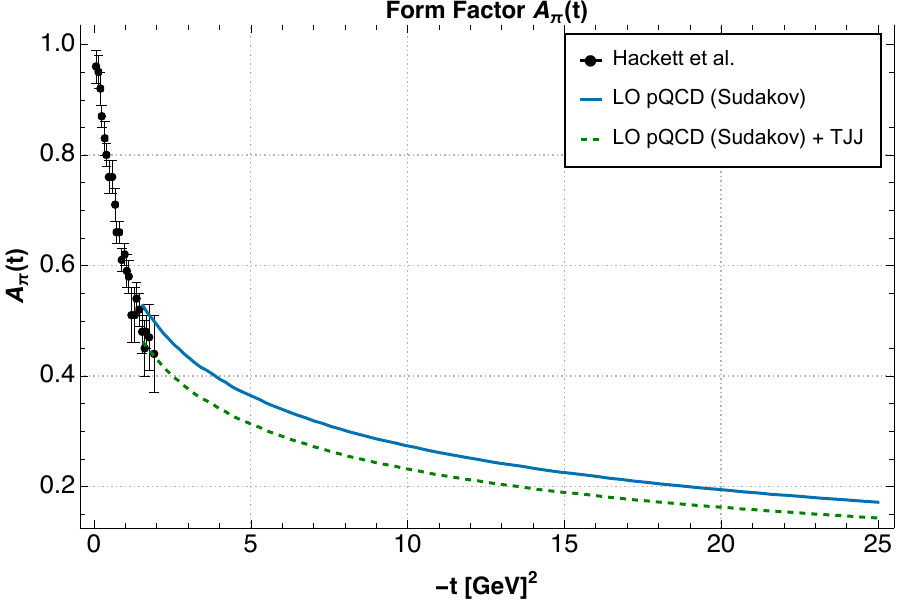}\hfill
\includegraphics[width=0.47\textwidth]{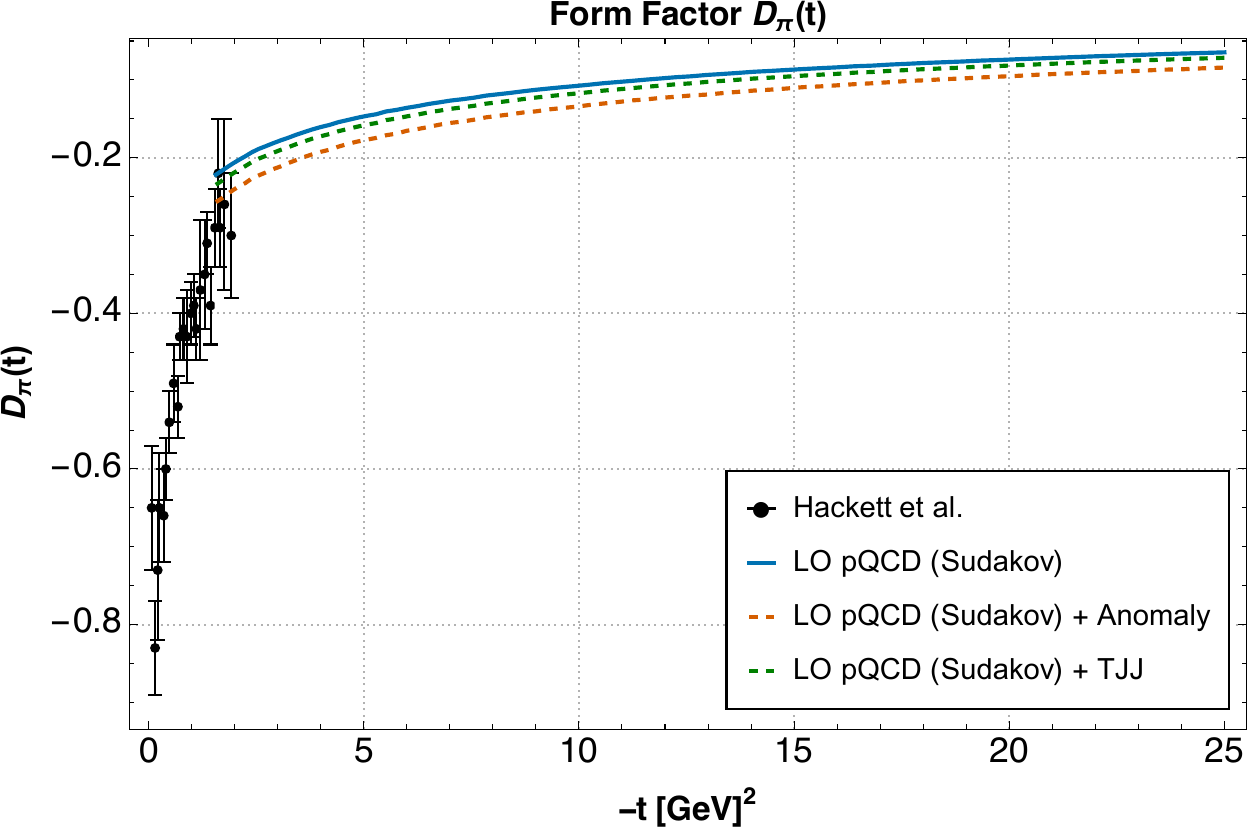}

\vspace{0.6ex}
\includegraphics[width=0.67\textwidth]{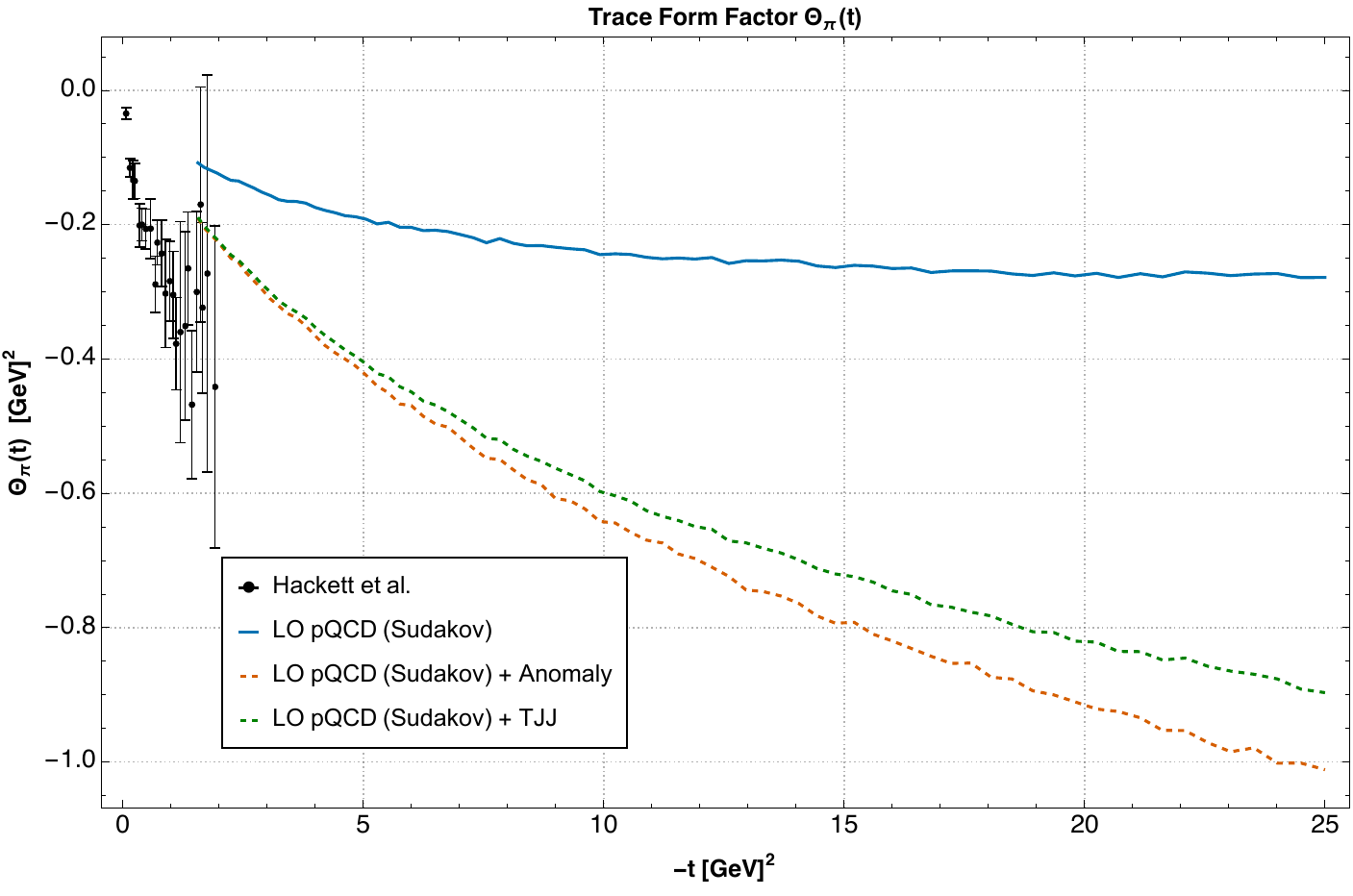}
\caption{Sudakov-resummation-improved pion GFFs obtained for $a_2(\mu_0)=0.2$, $\mu_0=1$ GeV, $m_q=0.33\,\, \text{GeV}$ and $\beta_\pi=0.24 \,\,\text{GeV}^{-1}$, compared with the lattice QCD data of Hackett et al.~\cite{Hackett:2023pyn}.  Top left: \(A_\pi(t)\), where the isolated anomaly cancels but the full \(TJJ\) insertion lowers the LO curve at small \(Q^2\).  Top right: \(D_\pi(t)\), where the anomaly gives an important contribution, while the plotted full \(TJJ\) correction is only part of the complete NLO hard kernel.  Bottom: \(\Theta_\pi(t)\), where the anomaly dominates the \(TJJ\) contribution.}
\label{fig:gff_comparison_letter}
\end{figure*}

The comparison with the lattice QCD results of Hackett et al.~\cite{Hackett:2023pyn} is shown in Fig.~\ref{fig:gff_comparison_letter}.  The curves should not be read as a complete low-energy model: at small \(-t\), chiral constraints, soft overlap, higher twist, and quark-mass terms in the trace must also be included.  The diagnostic point is instead the relative projection.  In \(A_\pi\) the isolated anomaly cancels, but the full \(TJJ\) correction lowers the LO curve and improves the local overlap with the data in the low-\(Q^2\) window.  In \(D_\pi\) the anomaly is sizeable, while the full displayed \(TJJ\) correction is more modest.  In \(\Theta_\pi\) the anomaly is the dominant part of the \(TJJ\) signal at both low and high momentum transfer.

The calculation is organized in three projections.  First, the LO kernel is matched onto \(A_\pi\), \(D_\pi\), and \(\Theta_\pi\).  Second, the full \(TJJ\) insertion is added.  Third, the part proportional to \(\mathcal{A}\) is tracked separately, so that the beta-function contribution can be compared with the full one-loop insertion.  This separation is essential: \(A_\pi\) tests the cancellation of the anomalous projection at leading twist, whereas \(D_\pi\) and especially \(\Theta_\pi\) retain the scalar anomaly component.

The three panels should therefore be read together.  The \(A_\pi\) plot is a test of the cancellation of the anomalous projection, not evidence that the full \(TJJ\) hard kernel is irrelevant.  The \(D_\pi\) plot shows the first sizeable scalar response, tied to the pressure-and-shear component encoded by the pion \(D\)-term.  The trace plot is the direct contraction of the EMT matrix element and gives the clearest signal.  This hierarchy is the main phenomenological output of the calculation.

The pattern also checks the tensor algebra.  The correction follows the projection expected from the \(TJJ\) tensor decomposition: it cancels in the isolated anomalous part of the momentum form factor, contributes to the \(D_\pi\) projection, and is enhanced in the trace observable.

\section{Interpretation}\label{interpretation}

The \(A_\pi\) correction deserves a precise formulation.  The part proportional to the anomaly coefficient cancels in the \(A_\pi\) projection and therefore does not contribute to the plotted anomaly curve.  This does not make the full \(TJJ\) insertion vanish: its non-anomalous tensor components shift the LO Sudakov result downward, improving the behaviour at low \(Q^2\) and giving a better local overlap with the lattice points in a small kinematic region. 

The \(D\)-term behaves differently because it is tied to the response of the system to spatial deformations.  In the spin-zero decomposition the same scalar structures that cancel from the anomalous part of \(A_\pi\) project more efficiently onto \(D_\pi\).  The anomaly gives an important contribution to this form factor, while the displayed full \(TJJ\) correction should be interpreted with care: it is not the complete NLO hard kernel.  The omitted non-anomalous terms could change the overlap with the lattice data also in the \(D_\pi\) channel.

The trace form factor gives the cleanest test.  Contracting the EMT matrix element with \(g_{\mu\nu}\) selects the scalar part directly, before assigning the result to the separate \(A_\pi\) and \(D_\pi\) tensor structures.  Here \(D_\pi\) is the form factor associated with the pressure-and-shear response of the pion.  In the trace channel the anomaly is the dominant contribution inside the \(TJJ\) correction both at low and at high momentum transfer.  

From Eq.~\eqref{eq:trace_explicit_letter}, \(\Theta_\pi\) contains an
	explicit factor of \(Q^2\) multiplying \(A_\pi\) and \(D_\pi\), which
	cancels the intrinsic \(1/Q^2\) falloff of both form factors at large
	momentum transfer.  As a result, the leading large-\(Q^2\) behavior of
	\(\Theta_\pi\) is controlled instead by the running coupling in Eq. \eqref{running},
	so that \(\Theta_\pi\) still approaches its expected asymptotic value
	of zero, but only logarithmically and only in the very high \(Q^2\)
	region.  In the kinematic window shown in Fig.~\ref{fig:gff_comparison_letter}, which is closer to the region
	accessible to near-term experiments and lattice studies, this slow
	logarithmic approach manifests itself instead as a
	flattening of the curves.
Treating the anomaly as a constrained nonperturbative input in the hard part therefore improves the trace prediction and, through the trace relation, also stabilizes the interpretation of \(D_\pi\).

The perturbative treatment adopted here is extended, in the numerical
comparison below, down to \(Q^2\sim 2\,\GeV^2\).  Its applicability in
this region relies on the Sudakov-resummation-improved factorization
framework \cite{Li:1992nu}, established for the
closely related case of the pion electromagnetic form factor.  
The approximation is expected to become increasingly
uncertain as \(Q^2\) decreases toward this region: Sudakov suppression
of the large-\(b\) region, while effective, is not absolute, and
residual sensitivity to the nonperturbative endpoint and large-\(b\)
behavior of the pion distribution amplitude introduces an uncertainty
that grows correspondingly.  The results below are displayed down to
\(Q^2\sim 2\,\GeV^2\) mainly to illustrate the qualitative trend of
the leading-order, anomaly, and full \(TJJ\) contributions across the
kinematic range covered by the lattice data, rather than as a claim
of quantitative precision in that specific region, which requires an analysis of the higher order corrections.

The comparison with lattice data is therefore qualitative but meaningful.  Current lattice points lie in a region where soft physics is still important, so one should not infer precision values for the anomaly contribution from the plots.  The useful information is the ordering of sensitivities.  A future calculation that includes soft-overlap terms and chiral constraints may change the normalization of the curves, but it should preserve the ordering if the trace-sector implementation is correct.

This pattern is physically meaningful because the trace contraction projects directly onto the anomaly-induced scalar structure, while the \(A_\pi\) and \(D_\pi\) projections combine the same \(TJJ\) insertion with different tensor components and with the pion convolution.  
The resulting hierarchy reflects the different sensitivity of the various form factors to this scalar part of the hard correction.

\section{Outlook}

This letter should be read as the perturbative trace-sector part of a broader description of pion gravitational form factors.  The momentum-space CFT analysis separates the transverse-traceless and trace sectors of stress-tensor correlators~\cite{Bzowski:2013sza}, while the anomaly-action and \(TJJ\) studies identify the scalar anomaly pole and its spectral interpretation~\cite{Giannotti:2008cv,Armillis:2009pq,Coriano:2018zdo}.  The recent GFF and sum-rule analyses embed the same structure in QCD factorization~\cite{Coriano:2024qbr,Coriano:2025sumrule}.  In the pion calculation this scalar component appears as one definite part of the one-loop \(TJJ\) insertion in the hard kernel.  The purpose of the present analysis is to isolate this part and to follow how it contributes to the separate \(A_\pi\), \(D_\pi\), and \(\Theta_\pi\) projections.

Several extensions are needed for a complete phenomenology.  First, the isolated \(TJJ\) insertion should be combined with the full NLO pion hard kernel, including the non-anomalous radiative terms and the evolution of the pion distribution amplitude.  Second, the finite-\(Q^2\) description should be matched to soft-overlap and chiral contributions, including the quark-mass term in the trace; this would also allow an explicit test of the sum rule of Eq.~\eqref{eq:sumrule_letter} within the hard-kernel calculation itself. The instanton-vacuum calculation of Ref.~\cite{Liu:2024vkj}, which includes twist-2 and twist-3 pion distribution amplitudes, semi-hard instanton effects, and pressure/shear distributions, provides a useful benchmark for this soft/semi-hard transition.  Future lattice data at larger \(-t\), with quark/gluon EMT separation and several pion masses, would directly test whether the hierarchy found here remains stable.

\section{Conclusions}

We have studied the contribution of the non-Abelian \(TJJ\) vertex to the pion gravitational form factors in QCD factorization.  The key result is that the beta-function-controlled scalar component of the hard kernel has different projections in the three channels: it cancels from the isolated anomalous part of \(A_\pi\), contributes appreciably to \(D_\pi\), and dominates the \(TJJ\) trace projection.

In the \(A_\pi\) channel the isolated anomalous contribution cancels, although the full \(TJJ\) insertion can still shift the leading Sudakov-improved curve through its non-anomalous tensor components.  In the \(D_\pi\) channel the anomaly gives an important scalar contribution, while a fully consistent comparison with data requires the remaining NLO hard terms.  The trace form factor \(\Theta_\pi\) is the cleanest projection: it selects the scalar sector directly and is therefore the most sensitive observable for the anomaly-induced component.\\
The Sudakov-resummation-improved treatment stabilizes the endpoint behaviour of the convolution and makes the channel hierarchy visible at finite momentum transfer.  The resulting pattern--cancellation in the anomalous part of \(A_\pi\), sizeable effects in \(D_\pi\), and dominance in the trace--is the main physical conclusion of the analysis. Would be interesting to investigate the relation between our analysis and the dilaton effective action approach detailed in \cite{Stegeman:2025tdl,Stegeman:2025sca}. More details of this investigation will be presented in forthcoming work.

\section*{Acknowledgements}

This work is partially supported by INFN under Iniziativa Specifica QG-sky and by National Science and Technology Council of the Republic of China under Grant No. NSTC-113-2112-M-001-024-MY3. C.C. thanks the Yang Institute and the Simons Center at Stony Brook for hospitality while completing this work. We dedicate this work to Prof. George Sterman.


\begin{thebibliography}{10}
\expandafter\ifx\csname url\endcsname\relax
  \def\url#1{\texttt{#1}}\fi
\expandafter\ifx\csname urlprefix\endcsname\relax\def\urlprefix{URL }\fi
\expandafter\ifx\csname href\endcsname\relax
  \def\href#1#2{#2} \def\path#1{#1}\fi

\bibitem{Ji:1996ek}
X.~Ji, {Gauge-Invariant Decomposition of Nucleon Spin}, Phys. Rev. Lett. 78
  (1997) 610--613.
\newblock \href {http://arxiv.org/abs/hep-ph/9603249}
  {\path{arXiv:hep-ph/9603249}}, \href
  {https://doi.org/10.1103/PhysRevLett.78.610}
  {\path{doi:10.1103/PhysRevLett.78.610}}.

\bibitem{Radyushkin:1996nd}
A.~V. Radyushkin, {Asymmetric gluon distributions and hard diffractive
  electroproduction}, Phys. Lett. B 385 (1996) 333--342.
\newblock \href {http://arxiv.org/abs/hep-ph/9604317}
  {\path{arXiv:hep-ph/9604317}}, \href
  {https://doi.org/10.1016/0370-2693(96)00528-X}
  {\path{doi:10.1016/0370-2693(96)00528-X}}.

\bibitem{Radyushkin:1996ru}
A.~V. Radyushkin, {Scaling Limit of Deeply Virtual Compton Scattering}, Phys.
  Lett. B 380 (1996) 417--425.
\newblock \href {http://arxiv.org/abs/hep-ph/9605431}
  {\path{arXiv:hep-ph/9605431}}, \href
  {https://doi.org/10.1016/0370-2693(96)00932-8}
  {\path{doi:10.1016/0370-2693(96)00932-8}}.

\bibitem{Vanderhaeghen:1998uc}
M.~Vanderhaeghen, P.~A.~M. Guichon, M.~Guidal, {Hard electroproduction of
  photons and mesons on the nucleon}, Phys. Rev. Lett. 80 (1998) 5064--5067.
\newblock \href {http://arxiv.org/abs/hep-ph/9806305}
  {\path{arXiv:hep-ph/9806305}}, \href
  {https://doi.org/10.1103/PhysRevLett.80.5064}
  {\path{doi:10.1103/PhysRevLett.80.5064}}.

\bibitem{Burkert:2018bqq}
V.~D. Burkert, L.~Elouadrhiri, F.~X. Girod, {The pressure distribution inside
  the proton}, Nature 557~(7705) (2018) 396--399.
\newblock \href {https://doi.org/10.1038/s41586-018-0060-z}
  {\path{doi:10.1038/s41586-018-0060-z}}.

\bibitem{Duran:2022xag}
B.~Duran, et~al., {Determining the gluonic gravitational form factors of the
  proton}, Nature 615~(7954) (2023) 813--816.
\newblock \href {http://arxiv.org/abs/2207.05212} {\path{arXiv:2207.05212}},
  \href {https://doi.org/10.1038/s41586-023-05730-4}
  {\path{doi:10.1038/s41586-023-05730-4}}.

\bibitem{Hackett:2023pyn}
D.~C. Hackett, P.~R. Oare, D.~A. Pefkou, P.~E. Shanahan, {Gravitational form
  factors of the pion from lattice QCD}, Phys. Rev. D 108~(11) (2023) 114504.
\newblock \href {http://arxiv.org/abs/2307.11707} {\path{arXiv:2307.11707}},
  \href {https://doi.org/10.1103/PhysRevD.108.114504}
  {\path{doi:10.1103/PhysRevD.108.114504}}.

\bibitem{Tong:2021ctu}
X.-B. Tong, J.-P. Ma, F.~Yuan, {Gluon gravitational form factors at large
  momentum transfer}, Phys. Lett. B 823 (2021) 136751.
\newblock \href {http://arxiv.org/abs/2101.02395} {\path{arXiv:2101.02395}},
  \href {https://doi.org/10.1016/j.physletb.2021.136751}
  {\path{doi:10.1016/j.physletb.2021.136751}}.

\bibitem{Tong:2022zax}
X.-B. Tong, J.-P. Ma, F.~Yuan, {Perturbative calculations of gravitational form
  factors at large momentum transfer}, JHEP 10 (2022) 046.
\newblock \href {http://arxiv.org/abs/2203.13493} {\path{arXiv:2203.13493}},
  \href {https://doi.org/10.1007/JHEP10(2022)046}
  {\path{doi:10.1007/JHEP10(2022)046}}.

\bibitem{Bzowski:2013sza}
A.~Bzowski, P.~McFadden, K.~Skenderis, {Implications of conformal invariance in
  momentum space}, JHEP 03 (2014) 111.
\newblock \href {http://arxiv.org/abs/1304.7760} {\path{arXiv:1304.7760}},
  \href {https://doi.org/10.1007/JHEP03(2014)111}
  {\path{doi:10.1007/JHEP03(2014)111}}.

\bibitem{Coriano:2024qbr}
C.~Corian\`o, S.~Lionetti, D.~Melle, R.~Tommasi, {The Gravitational Form
  Factors of Hadrons from CFT in Momentum Space and the Dilaton in Perturbative
  QCD}, Eur. Phys. J. C 85~(5) (2025) 498.
\newblock \href {http://arxiv.org/abs/2409.05609} {\path{arXiv:2409.05609}},
  \href {https://doi.org/10.1140/epjc/s10052-025-14104-1}
  {\path{doi:10.1140/epjc/s10052-025-14104-1}}.

\bibitem{Coriano:2024pionproton}
C.~Corian\`o, S.~Lionetti, D.~Melle, R.~Tommasi, {The Gravitational Form Factor
  of the Pion and Proton and the Conformal Anomaly}, EPJ Web Conf. 314 (2024)
  00030.
\newblock \href {http://arxiv.org/abs/2409.19586} {\path{arXiv:2409.19586}},
  \href {https://doi.org/10.1051/epjconf/202431400030}
  {\path{doi:10.1051/epjconf/202431400030}}.

\bibitem{Kumano:2017lhr}
S.~Kumano, Q.-T. Song, O.~V. Teryaev, {Hadron tomography by generalized
  distribution amplitudes in pion-pair production process $\gamma^* \gamma
  \rightarrow \pi^0 \pi^0 $ and gravitational form factors for pion}, Phys.
  Rev. D 97~(1) (2018) 014020.
\newblock \href {http://arxiv.org/abs/1711.08088} {\path{arXiv:1711.08088}},
  \href {https://doi.org/10.1103/PhysRevD.97.014020}
  {\path{doi:10.1103/PhysRevD.97.014020}}.

\bibitem{Savinov:2013hda}
V.~Savinov, {Measurement of $\gamma \gamma^{*} \to \pi^{0}$ transition form
  factor at Belle}, Nucl. Phys. B Proc. Suppl. 234 (2013) 287--290.
\newblock \href {https://doi.org/10.1016/j.nuclphysbps.2012.12.033}
  {\path{doi:10.1016/j.nuclphysbps.2012.12.033}}.

\bibitem{Belle:2015oin}
M.~Masuda, et~al., {Study of $\pi^0$ pair production in single-tag two-photon
  collisions}, Phys. Rev. D 93~(3) (2016) 032003.
\newblock \href {http://arxiv.org/abs/1508.06757} {\path{arXiv:1508.06757}},
  \href {https://doi.org/10.1103/PhysRevD.93.032003}
  {\path{doi:10.1103/PhysRevD.93.032003}}.

\bibitem{Efremov:1979qk}
A.~V. Efremov, A.~V. Radyushkin, Factorization and asymptotical behavior of
  pion form factor in qcd, Phys. Lett. B 94 (1980) 245.
\newblock \href {https://doi.org/10.1016/0370-2693(80)90869-2}
  {\path{doi:10.1016/0370-2693(80)90869-2}}.

\bibitem{Lepage:1980fj}
G.~P. Lepage, S.~J. Brodsky, Exclusive processes in perturbative quantum
  chromodynamics, Phys. Rev. D 22 (1980) 2157.
\newblock \href {https://doi.org/10.1103/PhysRevD.22.2157}
  {\path{doi:10.1103/PhysRevD.22.2157}}.

\bibitem{Coriano:2025sumrule}
C.~Corian\`o, S.~Lionetti, D.~Melle, L.~Torcellini, {A dilaton sum rule for the
  conformal anomaly form factor in QCD at order $\alpha_s$}, Eur. Phys. J. C
  85~(9) (2025) 983.
\newblock \href {http://arxiv.org/abs/2504.01904} {\path{arXiv:2504.01904}},
  \href {https://doi.org/10.1140/epjc/s10052-025-14686-w}
  {\path{doi:10.1140/epjc/s10052-025-14686-w}}.

\bibitem{Giannotti:2008cv}
M.~Giannotti, E.~Mottola, {The Trace Anomaly and Massless Scalar Degrees of
  Freedom in Gravity}, Phys. Rev. D79 (2009) 045014.
\newblock \href {http://arxiv.org/abs/0812.0351} {\path{arXiv:0812.0351}},
  \href {https://doi.org/10.1103/PhysRevD.79.045014}
  {\path{doi:10.1103/PhysRevD.79.045014}}.

\bibitem{Armillis:2009pq}
R.~Armillis, C.~Corian\`{o}, L.~Delle~Rose, {Conformal Anomalies and the
  Gravitational Effective Action: The $TJJ$ Correlator for a Dirac Fermion},
  Phys. Rev. D81 (2010) 085001.
\newblock \href {http://arxiv.org/abs/0910.3381} {\path{arXiv:0910.3381}},
  \href {https://doi.org/10.1103/PhysRevD.81.085001}
  {\path{doi:10.1103/PhysRevD.81.085001}}.

\bibitem{Armillis:2010qk}
R.~Armillis, C.~Corian\`o, L.~Delle~Rose, {Trace Anomaly, Massless Scalars and
  the Gravitational Coupling of QCD}, Phys. Rev. D82 (2010) 064023.
\newblock \href {http://arxiv.org/abs/1005.4173} {\path{arXiv:1005.4173}},
  \href {https://doi.org/10.1103/PhysRevD.82.064023}
  {\path{doi:10.1103/PhysRevD.82.064023}}.

\bibitem{Coriano:2012wp}
C.~Coriano, L.~Delle~Rose, E.~Mottola, M.~Serino, {Graviton Vertices and the
  Mapping of Anomalous Correlators to Momentum Space for a General Conformal
  Field Theory}, JHEP 08 (2012) 147.
\newblock \href {http://arxiv.org/abs/1203.1339} {\path{arXiv:1203.1339}},
  \href {https://doi.org/10.1007/JHEP08(2012)147}
  {\path{doi:10.1007/JHEP08(2012)147}}.

\bibitem{Coriano:2014gja}
C.~Corian\`o, A.~Costantini, L.~Delle~Rose, M.~Serino, {Superconformal sum
  rules and the spectral density flow of the composite dilaton (ADD) multiplet
  in $\mathcal{N}=1$ theories}, JHEP 06 (2014) 136.
\newblock \href {http://arxiv.org/abs/1402.6369} {\path{arXiv:1402.6369}},
  \href {https://doi.org/10.1007/JHEP06(2014)136}
  {\path{doi:10.1007/JHEP06(2014)136}}.

\bibitem{Coriano:2018zdo}
C.~Corian\`o, M.~M. Maglio, {Renormalization, Conformal Ward Identities and the
  Origin of a Conformal Anomaly Pole}, Phys. Lett. B781 (2018) 283--289.
\newblock \href {http://arxiv.org/abs/1802.01501} {\path{arXiv:1802.01501}},
  \href {https://doi.org/10.1016/j.physletb.2018.04.003}
  {\path{doi:10.1016/j.physletb.2018.04.003}}.

\bibitem{Coriano:2020ees}
C.~Corian\`o, M.~M. Maglio, {Conformal field theory in momentum space and
  anomaly actions in gravity: The analysis of three- and four-point function},
  Phys. Rept. 952 (2022) 2198.
\newblock \href {http://arxiv.org/abs/2005.06873} {\path{arXiv:2005.06873}},
  \href {https://doi.org/10.1016/j.physrep.2021.11.005}
  {\path{doi:10.1016/j.physrep.2021.11.005}}.

\bibitem{Coriano:2018bbe}
C.~Corian\`o, M.~M. Maglio, {Exact Correlators from Conformal Ward Identities
  in Momentum Space and the Perturbative $TJJ$ Vertex}, Nucl. Phys. B938 (2019)
  440--522.
\newblock \href {http://arxiv.org/abs/1802.07675} {\path{arXiv:1802.07675}},
  \href {https://doi.org/10.1016/j.nuclphysb.2018.11.016}
  {\path{doi:10.1016/j.nuclphysb.2018.11.016}}.

\bibitem{Coriano:2026vya}
C.~Corian\`o, S.~Lionetti, D.~Melle, L.~Torcellini, {Anomaly-mediated Scalar
  Gravitational Interactions and the Coupling of Conformal Sectors}, arXiv
  preprint (3 2026).
\newblock \href {http://arxiv.org/abs/2603.28966} {\path{arXiv:2603.28966}}.

\bibitem{2016arXiv160609220M}
E.~{Mottola}, {Scalar Gravitational Waves in the Effective Theory of Gravity},
  ArXiv e-prints (Jun. 2016).
\newblock \href {http://arxiv.org/abs/1606.09220} {\path{arXiv:1606.09220}}.

\bibitem{Botts:1989kf}
J.~Botts, G.~Sterman, Hard elastic scattering in qcd: Leading behavior, Nucl.
  Phys. B 325 (1989) 62.
\newblock \href {https://doi.org/10.1016/0550-3213(89)90372-6}
  {\path{doi:10.1016/0550-3213(89)90372-6}}.

\bibitem{Li:1992nu}
H.-n. Li, G.~Sterman, The perturbative pion form factor with sudakov
  suppression, Nucl. Phys. B 381 (1992) 129.
\newblock \href {https://doi.org/10.1016/0550-3213(92)90643-P}
  {\path{doi:10.1016/0550-3213(92)90643-P}}.

\bibitem{Sterman:1986aj}
G.~Sterman, Summation of large corrections to short distance hadronic
  cross-sections, Nucl. Phys. B 281 (1987) 310.
\newblock \href {https://doi.org/10.1016/0550-3213(87)90258-6}
  {\path{doi:10.1016/0550-3213(87)90258-6}}.

\bibitem{Liu:2024vkj}
W.-Y.~Liu, E.~Shuryak and I.~Zahed, Pion gravitational form factors in the
  QCD instanton vacuum. II, Phys. Rev. D 110 (2024) 054022.
\newblock \href {https://doi.org/10.1103/PhysRevD.110.054022}
  {\path{doi:10.1103/PhysRevD.110.054022}}.
  
\bibitem{Stegeman:2025tdl}
R.~Stegeman and R.~Zwicky, Gluon gravitational D-form factor: the
  $\sigma$-meson as a dilaton confronted with lattice data II, JHEP 05
  (2026) 159.
\newblock \href {https://doi.org/10.1007/JHEP05(2026)159}
  {\path{doi:10.1007/JHEP05(2026)159}}.

\bibitem{Stegeman:2025sca}
R.~Stegeman and R.~Zwicky, Gravitational D-form factor: the
  $\sigma$-meson as a dilaton confronted with lattice QCD data I, JHEP 03
  (2026) 184.
\newblock \href {https://doi.org/10.1007/JHEP03(2026)184}
  {\path{doi:10.1007/JHEP03(2026)184}}.

\end{thebibliography}
\end{document}